\documentclass[showpacs,preprint]{revtex4}
\usepackage{graphicx}
\usepackage{dcolumn}
\usepackage{amsmath}
\usepackage[latin1]{inputenc}
\usepackage{graphicx, psfrag}
\usepackage{amssymb}
\usepackage[colorlinks=true, citecolor=blue, urlcolor = blue, linkcolor= red, bookmarks=true]{hyperref}
\usepackage{float}
\usepackage{amsmath}
\usepackage{amsfonts}
\usepackage{dcolumn}
\usepackage{hyperref}
\usepackage{subfigure}
\usepackage{epstopdf}
\usepackage{booktabs}
\usepackage{multirow}

\makeatletter
\def\btt#1{\texttt{\@backslashchar#1}}
\DeclareRobustCommand\bblash{\btt{\@backslashchar}} \makeatother

\makeatletter
\def\btt#1{\texttt{\@backslashchar#1}}
\DeclareRobustCommand\bblash{\btt{\@backslashchar}} \makeatother

\begin{document}
\title{ Black Hole Evaporation Process and  Tangherlini-Reissner-Nordstr\"om Black Holes Shadow }
\author{Balendra Pratap Singh$^{a}$}\email{balendrap.singh@ddn.upes.ac.in}

\affiliation{ $^{a}$ Department of Physics, Applied Science Cluster, SOAE, UPES, Energy Acres, Bidholi, Via Prem Nagar,  Dehradun, Uttarakhand, 248007, India}

\begin{abstract}
In this article, we study the black hole evaporation process and shadow property of the Tangherlini-Reissner-Nordstr\"om (TRN) black holes. The TRN black holes are the higher-dimensional extension of the Reissner-Nordstr\"om (RN) black holes and are characterized by their mass $M$, charge $q$, and spacetime dimensions $D$.  In higher-dimensional spacetime, the black hole evaporation occurs rapidly, causing the black hole's horizon to shrink. We derive the rate of mass loss for the higher-dimensional charged black hole and investigate the effect of higher-dimensional spacetime on charged black hole shadow.  We derive the complete geodesic equations of motion with the effect of spacetime dimensions $D$. We determine impact parameters by maximizing the black hole's effective potential and estimate the critical radius of photon orbits. 
The photon orbits around the black hole shrink with the effect of the increasing number of spacetime dimensions. To visualize the shadows of the black hole, we derive the celestial coordinates in terms of the black hole parameters.  We use the observed results of M87 and Sgr A$^{*}$ black hole from the Event Horizon Telescope and estimate the angular diameter of the charge black hole shadow in the higher-dimensional spacetime. We also estimate the energy emission rate of the black hole. Our finding shows that the angular diameter of the black hole shadow decreases with the increasing number of spacetime dimensions $D$.
 
\end{abstract}
\maketitle
 {\section{Introduction}}
The higher-dimensional spacetimes are the extension of our four-dimensional world. The theories of gravity and their properties get modified in the higher-dimensional spacetime. After the discovery of general relativity, Kaluza \cite{Kaluza:1921tu} and Klein \cite{Klein:1926tv} added another extra dimension to Einstein's general theory of relativity.  Examining black holes in higher-dimensional spacetime is intriguing due to their peculiarity. String theory is the theoretical framework that supports the study of black holes in higher dimensions \cite{Emparan:2008eg}.  There is a chance of producing higher-dimensional micro black holes during particle collisions in the Large Hadron Collider at TeV scale \cite{Dimopoulos:2001hw, Seitova:2019adv}. 
The equations of motion of particles around the black hole are significantly altered in higher dimensions \cite{Sharif:2016rhs, Dadhich:2021vdd}. The spectroscopic study of hydrogen atoms in the higher-dimensional spacetime commits the existence of the extra number of spacetime dimensions \cite{Burgbacher:1999sha, Caruso:2012daf, Shaqqor:2009cha, Zhou:2014xbw, Luo:2006ck, Luo:2006ad}. Maxwell's electrodynamics equations also get modified in the extra-dimensional spacetime  \cite{Frolov:2021tqe}. In this context, the study of higher-dimensional black holes has attracted the attention of various researchers for decades. In 1963, Tangherlini initiated the study of black holes in higher-dimensional spacetime by extending the Schwarzschild solution \cite{Tangherlini:1963bw}. In 1986, Myers and Perry found the exact solution for higher-dimensional rotating spacetime \cite{Myers:1986un}.  The properties of black holes are strongly affected by the spacetime dimensions and have been extensively researched \cite{Ishibashi:2003ap, Ghosh:2008zza, Ghosh:2001pv, Ghosh, Ghosh:2001fb, Cai:2020igv, Andre:2021ctu, Ma:2020jls, Paul:2023gzj, Belhaj:2015hha, Uniyal:2022xnq, Nashed:2018piz, Kunz:2013jja, Frassino:2022zaz, Pourhassan:2017kmm, Zhao:2022dgc}.

Hawking posited that a black hole emits particles like neutrinos and photons, known as Hawking radiation. \cite{Hawking:1974rv, Hawking:1975vcx}.  The measurements of these particles can be done by including the quantum gravitational effects. With the help of Hawking radiation people are finding the relations between general relativity and quantum mechanics \cite{Vaz:2002xb}.  A black hole emits thermal radiation, causing it to lose its hairs such as mass, charge, and spin. The rate of change of the mass of a black hole depends on the surface gravity of the black hole. The black hole keeps losing its mass until it completely evaporates. The lifetime of a four-dimensional Schwarzschild black hole is $t\approx M_{0}^3$, where $M_{0}$ is the initial mass of the black hole \cite{Page:1976df}. The static-charged black hole loses its charge as well as its mass, and over time the charged black hole evolves into the static black hole \cite{Hiscock:1990ex}.
The four-dimensional rotating black hole loses its angular momentum faster than its mass and with time it evolves toward the Schwarzschild black hole \cite{Page:1976df,Taylor:1998dk}. In higher-dimensional spacetime, the black hole horizon shrinks and this increases the surface gravity which causes the emission rate to fast \cite{Xu:2020xsl}, and hence a black hole in higher dimensions evaporates faster than the four-dimensional spacetime \cite{Xu:2019wak}.

The physical appearance of the black hole comes in nature due to its shadow property.  Black holes, although dark, can be visualized by their shadows. The black hole shadow is a dark disc surrounded by bright photon rings. 
Event Horizon Telescope (EHT) collaborators published the observed shadow images of the M87 black hole in \cite{EventHorizonTelescope:2019dse, EventHorizonTelescope:2019uob, EventHorizonTelescope:2019jan, EventHorizonTelescope:2019ths, EventHorizonTelescope:2019pgp, EventHorizonTelescope:2019ggy} and the Sgr A$^{*}$ black hole in \cite{EventHorizonTelescope:2021bee, EventHorizonTelescope:2021srq, EventHorizonTelescope:2022wkp, EventHorizonTelescope:2022apq, EventHorizonTelescope:2022wok, EventHorizonTelescope:2022exc, EventHorizonTelescope:2022urf, EventHorizonTelescope:2022xqj}. The study of black hole shadow in higher-dimensional spacetime has been the subject of interest for the last few years. Papnoi et al. investigate the shadow property of 5$D$ rotating Myers-Perry black holes \cite{Papnoi:2014aaa}. The shadow property for Gauss-Bonnet gravity rotating black holes in six-dimensional spacetime has been done by  \cite{Abdujabbarov:2015rqa}.  Amir et al. investigate the shadow properties of five-dimensional EMCS black holes in \cite{Amir:2017slq}.
Belhaj et al. extended the study of the quintessential black hole shadow in arbitrary dimensions \cite{Belhaj:2020rdb}. The authors of \cite{Nozari:2023flq} investigate the shadow properties of AdS higher-dimensional black holes of Einstein-Horndeski- Maxwell gravity. Banerjee et al. study the extra-dimensional spacetime using EHT observations in \cite{Banerjee:2022jog}. 

In our previous work, we highlight the shadow properties of Schwarzschild's black hole in higher dimensions 
\cite{Singh:2017vfr} and its rotating counterpart \cite{Singh:2023ops}.  Our previous finding shows that in higher dimensional spacetime the size of black hole shadow decreases with the increasing number of spacetime dimensions. In this work, we study the cause of this effect. This study aims to investigate the process of black hole evaporation, generalize shadow properties for the TRN class of black holes, and determine their angular diameter.  Generalizing the null geodesic equations of motion for photons into the arbitrary number of spacetime dimensions, we find the effective potential of the black hole.  We maximize the effective potential to obtain photon orbits around the charged black hole in higher dimensions. The spacetime dimensions significantly affect the charged black hole shadow. In our study, we show that the effective size and the angular diameter of the black hole shadow decrease with the increasing number of spacetime dimensions, and also the effect of the charge is dominated by the increasing number of spacetime dimensions.

This paper is arranged in the following manner: In section two  (\ref{sect2}),  we describe the black hole metric and its thermodynamical properties. We derive the complete null geodesic equations in section three (\ref{sect3}). We calculate the equation of rate of mass loss in (\ref{sect4}) and in section  (\ref{sect5}), we study the black hole shadow with the increasing number of spacetime dimensions. We check the consistency of the higher-dimensional black hole with the M87 and Sgr A$^*$ black hole shadow in (\ref{sect6}). We calculate the energy emission rate of the higher-dimensional charged black hole in  (\ref{sect7}), and finally,  we conclude our results in section (\ref{sect8}).
\section{General formalism of the black hole metric}\label{sect2}
The action  for an asymptotically flat higher-dimensional charged black hole can be written as
\begin{equation}
    S = \frac{1}{16 \pi G} \int_{M} d^{d}x \sqrt{-g}(R - F^{\mu\nu}F_{\mu\nu}),
\end{equation}
where $R$ is the Ricci scalar and $F_{\mu\nu}$ is the Maxwell tensor. The spherically symmetric charged black hole metric in $D$-dimensional spacetime with natural units $(G = c = \hbar  =  1)$ is given by \cite{Tangherlini-1963,Aman:2005xk, Destounis:2019hca, Gao:2008jy, Xu:2019wak} 
\begin{equation} \label{metric}
ds^2=-\left(1-{\frac{\mu}{r^{D-3}}+\frac{\nu^2}{r^{2(D-3)}}}\right)dt^{2} + \left(1-{\frac{\mu}{r^{D-3}}+\frac{\nu^2}{r^{2(D-3)}}}\right)^{-1}dr^{2} + r^2d\Omega^{2}_{(D-2)}
\end{equation}
where
\begin{equation}
  d\Omega^2_{D-2}=d{\theta_1}^2+{\sin^2{\theta_1}}{d{\theta_2}^2}+, . . . ,+\prod_{i=1}^{D-3}\sin^2\theta_id{\theta}_{D-2}^2,
\end{equation}
is the line element on $(D-2)$-dimensional unit sphere \cite{Yang:2018cim}, and 
 \begin{equation}
\mu=\frac{16{\pi}M}{(D-2)\Omega_{D-2}},
\end{equation}
with
\begin{equation}
	\Omega_{D-2}=\frac{2\pi^{\frac{D-1}{2}}}{\Gamma\left(\frac{D-1}{2}\right)},
\end{equation}
and the parameter $\nu$ is associated with black hole charge $q$ \cite{Yang:2018cim}  via 
\begin{equation}
\nu = \sqrt{\frac{8 \pi}{{ (D-2)(D-3)}}}\frac{q}{\Omega_{D-2}}.
\end{equation}
The parameters $\mu$ and $\nu$ are the ADM mass and electric charge of the black hole respectively \cite{Myers:1986un}. The black hole metric (\ref{metric}) is the higher-dimensional extension of the RN black hole solution. In the absence of the charge parameter $q = 0$, the black hole metric (\ref{metric}) reduces to the Schwarzschild-Tangherlini black hole \cite{Singh:2017vfr}. In four-dimensional spacetime, the black hole mass is $2M$ and, the black hole charge is $q$, which is the RN black hole \cite{Chandrasekhar:1992} and for $q=0$ it reduces to the Schwarzschild black hole \cite{Chandrasekhar:1992}.  One can find the roots of the black hole metric (\ref{metric}) by simply solving $f(r) = 0$, as
\begin{equation} \label{horizon}
r_{\pm} = \left(\frac{\mu}{2} \pm \frac{\mu}{2}\sqrt{1 - \frac{4\nu^2}{\mu^2}}\right)^{1/(D-3)},
\end{equation}
where $r_{+}$ corresponds to the outer horizon and $r_{-}$ is the inner or Cauchy horizon of the black hole. Through horizon equation (\ref{horizon}), one can find
\begin{equation}
    r_{+}^{D-3} + r_{-}^{D-3} = \mu \quad \quad \text{and} \quad \quad r_{+}^{D-3} r_{-}^{D-3} = {\nu}^2.
\end{equation}
The TRN black hole solution develops naked singularity when $\nu^2 > \mu^2/4$ with the central singularity  $r = 0$. The area of the event horizon of the TRN black hole is given by \cite{Aman:2005xk}:
\begin{equation}
    A = \Omega_{({D-2})} r_{+}^{(D-2)}.
\end{equation}
In this context, \cite{Falcke:2003} the entropy of the black hole follows the given expression
\begin{equation}
    S = \frac{1}{4}\Omega_{(D-2)}r_{+}^{(D-2)},
\end{equation}
here the entropy of the black hole varies with the black hole horizon and the spacetime dimension $D$. The temperature of the black hole depends on the surface gravity at the horizon of the black hole and is obtained as
\begin{equation}
T=\frac{(D-3)\left(M^2-Q^2+M\sqrt{M^2-Q^2}\right)}{2\pi\left(M+\sqrt{M^2-Q^2}\right)^{\frac{2D-5}{D-3}}}.
\label{temperature}
\end{equation}
As we go to the higher-dimensional spacetime the horizon of the black hole decreases and the temperature of the black hole increases.

\section{ Basic Equations for Null Geodesics} \label{sect3}
In this section, we derive the null geodesic equations of motion to find the orbits of photons around the TRN black holes. The required equations of motion can be obtained with the help of the associated Lagrangian function. The Lagrangian of the black hole takes the following form \cite{Chandrasekhar:1992} 
\begin{equation}
		\mathcal{L} = \frac{1}{2} g_{\mu \nu}\dot{x}^{\mu}\dot{x}^{\nu},
\end{equation}
where $g_{\mu\nu}$ is the metric tensor of the black hole which can be found via the black hole metric Eq.~$(\ref{metric})$,  and over dot represents the derivative with respect to the affine parameter $\tau$.  We calculate the canonically conjugate momentum via the Lagrangian function of the black hole as \cite{Chandrasekhar:1992} 
\begin{eqnarray}
&& P_{t}=\left(1-{\frac{\mu}{r^{D-3}}+\frac{\nu^2}{r^{2(D-3)}}}\right)\dot{t}=\mathcal{E}, \label{p1}\\
&&{P_r}= \left(1-{\frac{\mu}{r^{D-3}}+\frac{\nu^2}{r^{2(D-3)}}}\right)^{-1}\dot{r}, \label{p3}\\
&&P_{\theta_i}=r^2\sum_{i=1}^{D-3}\prod_{n=1}^{i-1}\sin^2\theta_n\dot
{\theta}_{D-3}, \quad \:\:\:  i=1, . . . ,D-3 \label{p4}\\
&& P_{\phi}=r^2\prod_{i=1}^{D-3}\sin^2\theta_i\dot{\theta}_{D-2}=L, \label{p2}
\end{eqnarray}
where $P_\phi=P_{\theta_{D-2}}$ and the parameter $\mathcal{E}$  and $L$ represent the energy and the angular momentum respectively. In four-dimensional spacetime, the above momentum equations (\ref{p1}-\ref{p2}) reduce for the RN black hole and take the form
\begin{eqnarray}
&&P_{\theta_{1}}=r^2\dot{\theta_{1}},\nonumber\\ 
&&P_{\phi}=r^2\sin^2\theta_{1}\dot{\theta_{2}}.\nonumber
\end{eqnarray}
Next, we apply the Hamilton-Jacobi method to obtain our complete equations of motion. In higher-dimensional spacetime, the Hamilton-Jacobi equation reads \cite{Chandrasekhar:1992, Frolov:1998wf} 
\begin{equation}
\frac{\partial S}{\partial {\tau}}  =\mathcal{H}=-\frac{1}{2}  g^{\mu\nu}\frac{\partial S}{\partial 		x^\mu}\frac{\partial S}{\partial x^\nu}, \label{j}
\end{equation} 
where $S$ represents the Jacobian action, and $g^{\mu\nu}$ is the inverse of the metric tensor. Using black hole metric $(\ref{metric})$  in Eq.~$(\ref{j})$, one can obtain
\begin{eqnarray}\label{action}
-2\frac{\partial S}{\partial {\tau}} &=& -\frac{1}{\left(1-{\frac{\mu}{r^{D-3}}+\frac{\nu^2}{r^{2(D-3)}}}\right)}\left(\frac{\partial S_t}{\partial t}\right)^2+ \left(1-{\frac{\mu}{r^{D-3}}+\frac{\nu^2}{r^{2(D-3)}}}\right)\left( \frac{\partial S_r}{\partial r}\right)^2 \nonumber \\
&+& 
\sum_{i=1}^{D-3}\frac{1}{r^2\prod\limits_{n=1}^{i-1}\sin^2\theta_n}\left(\frac{\partial S_{\theta_i}}{\partial {\theta}_i }\right)^2
+ \frac{1}{r^2\prod\limits_{i=1}^{D-3}\sin^2\theta_i}\left(\frac{\partial S_{\phi}}{\partial {\phi}}\right)^2.
\end{eqnarray}
The above Eq.~($\ref{action}$) is the higher-dimensional partial differential equations with a set of coordinates. To separate these coordinates, we choose an additive separable solution \cite{Frolov:1998wf}
\begin{equation}\label{sep}
S= \frac{1}{2}{m^2} \tau - \mathcal{E}t+  {L}\phi +
S_r (r) + \sum\limits_{i=1}^{D-3} S_{\theta_i}(\theta_i),
\end{equation}
\begin{figure}
    \centering
    \includegraphics[scale = 0.6]{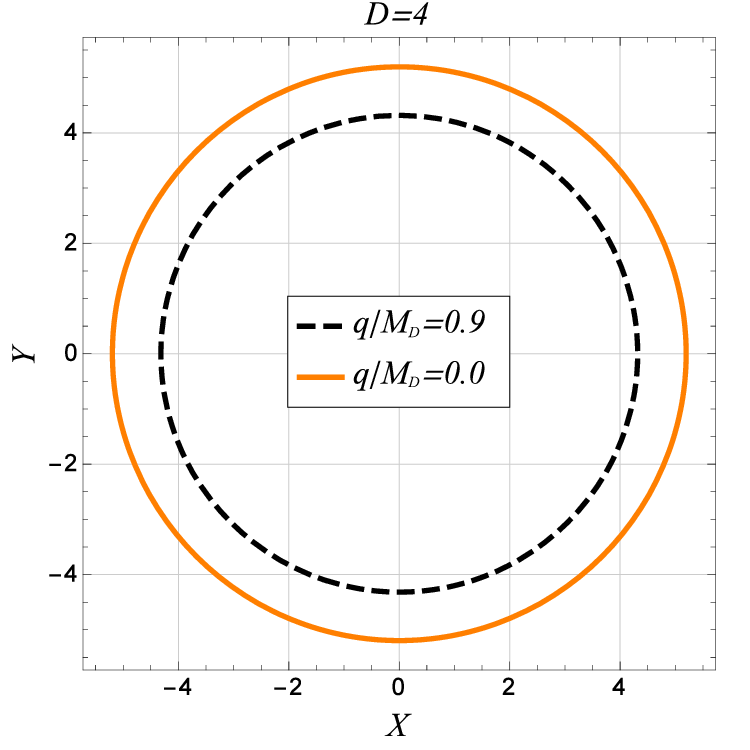}
    \includegraphics[scale = 0.6]{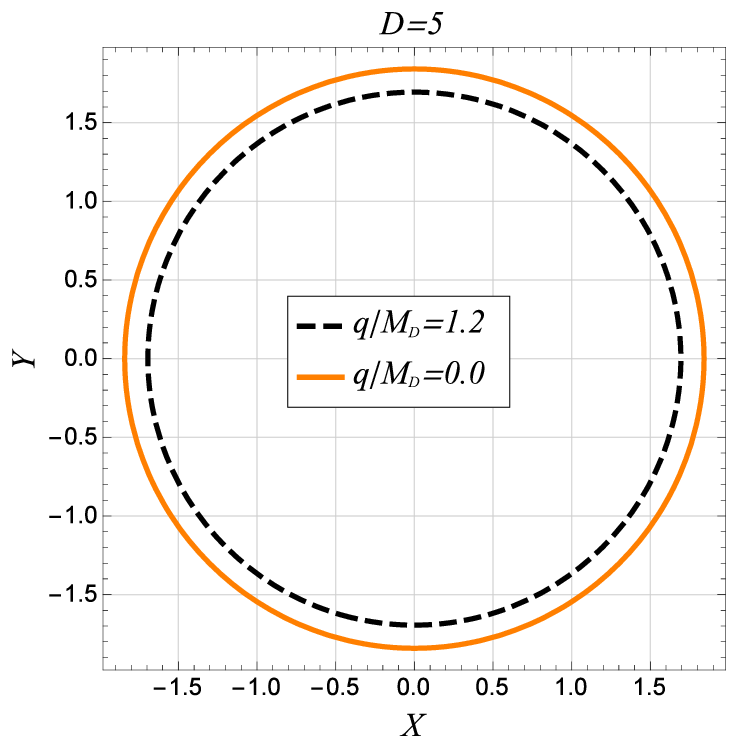}\\
    \includegraphics[scale = 0.6]{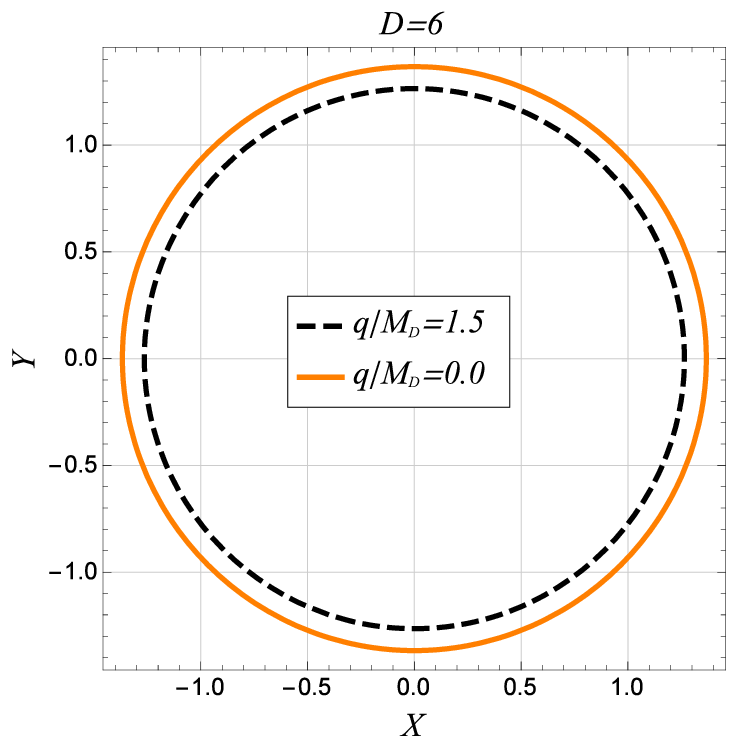}
    \includegraphics[scale = 0.6]{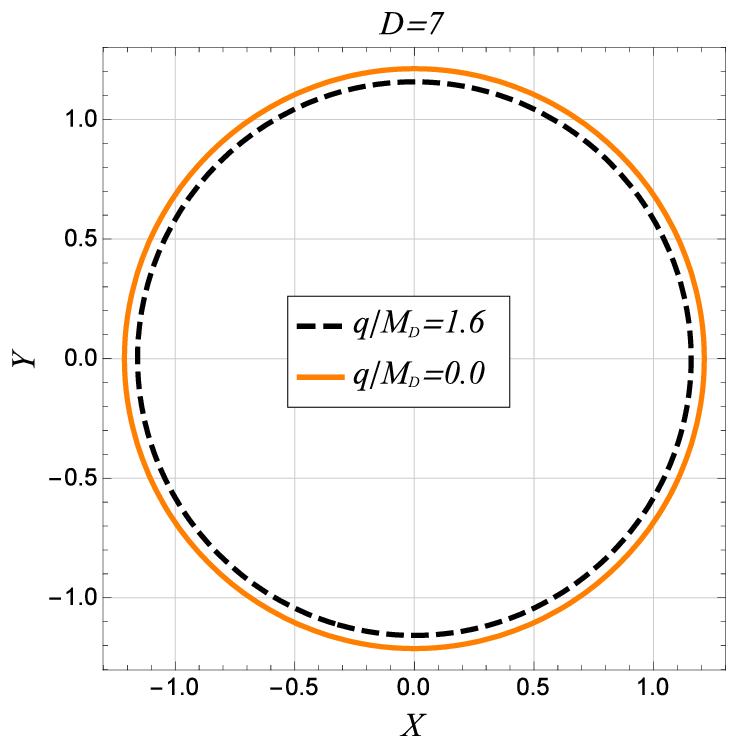}\\
    \caption{Plot showing the variation of black hole shadow with charge $q$ and spacetime dimensions $D$.}
    \label{figure1}
\end{figure}
\begin{table}[]
\begin{center}
\begin{tabular}{|c|c|c|c|c|c|c|c|}
\hline
\multicolumn{1}{|l|}{\textbf{$D$}} & \multicolumn{1}{l|}{\textbf{$q/M_D$}} & \textbf{$r_c/M_D$} & \multicolumn{1}{l|}{\textbf{$\eta + \xi^2$}} & \textbf{$D$} & \textbf{$q/M_D$} & \textbf{$r_c/M_D$} & \textbf{$\eta + \xi^2$} \\ \hline
\multirow{4}{*}{4} & 0.0 & 3.0000 & 27.0000 & \multirow{4}{*}{6} & 0.0 & 1.0607 & 1.8754 \\ \cline{2-4} \cline{6-8} 
 & 0.6 & 2.7369 & 23.6069 &  & 0.6 & 1.0464 & 1.8443 \\ \cline{2-4} \cline{6-8} 
 & 0.9 & 2.2937 & 18.6557 &  & 0.9 & 1.0258 & 1.8015 \\ \cline{2-4} \cline{6-8} 
 & 1.0 & 2.0000 & 16.0000 &  & 1.5 & 0.8947 & 1.6009 \\ \hline
\multirow{4}{*}{5} & 0.0 & 1.2917 & 3.3963 & \multirow{4}{*}{7} & 0.0 & 0.9762 & 1.3341 \\ \cline{2-4} \cline{6-8} 
 & 0.6 & 1.2676 & 3.2754 &  & 0.6 & 0.9708 & 1.3257 \\ \cline{2-4} \cline{6-8} 
 & 0.9 & 1.2145 & 3.1064 &  & 0.9 & 0.9634 & 1.3144 \\ \cline{2-4} \cline{6-8} 
 & 1.1 & 1.1526 & 2.9282 &  & 1.6 & 0.8841 & 1.2252 \\ \hline
\end{tabular}
\caption{\label{table1} Variation of critical radius $r_{c}$ and $\eta + \xi^2$ with the black hole charge $q$ and spacetime dimension $D$ .}
\end{center}
\end{table}
where $m$ is the mass of the test particle, which is zero for the case of a photon, $S_r(r)$ is the function of radial coordinate, and  $S_{\theta_i}(\theta_i)$  is the function of angular coordinate.  In the above Eq.~(\ref{sep}), the second term $\mathcal{E}t$ corresponds to the energy conservation, and the third term ${L}\phi$ corresponds to the angular momentum conservation. Using the separable solution (\ref{sep}) in the Hamilton-Jacobi Eq.~(\ref{j}), we obtain separate integrable equations for the radial and the angular coordinates as \cite{Frolov:1998wf}
\begin{eqnarray}
r^4\left(1-{\frac{\mu}{r^{D-3}}+\frac{\nu^2}{r^{2(D-3)}}}\right)^2\left(\frac{\partial S_r}{\partial r}\right)^2 &=&  {\mathcal{E}^2} {r^4} - r^2\left(1-{\frac{\mu}{r^{D-3}}+\frac{\nu^2}{r^{2(D-3)}}}\right)\left(\mathcal{K}+{{L}}^2\right),\label{sr} \\
{\displaystyle\sum_{i=1}^{D-3}\frac{1}{\displaystyle\prod_{n=1}^{i-1}\sin^2\theta_i}}\left(\frac{\partial S_{\theta_i}}{\partial \theta_i}\right)^2 &=& \mathcal{K}-\prod_{i=1}^{D-3}{{L}}^2\cot^2\theta_i, \label{stheta}
\end{eqnarray}

\begin{figure}
    \centering
    \includegraphics[width=0.7\linewidth]{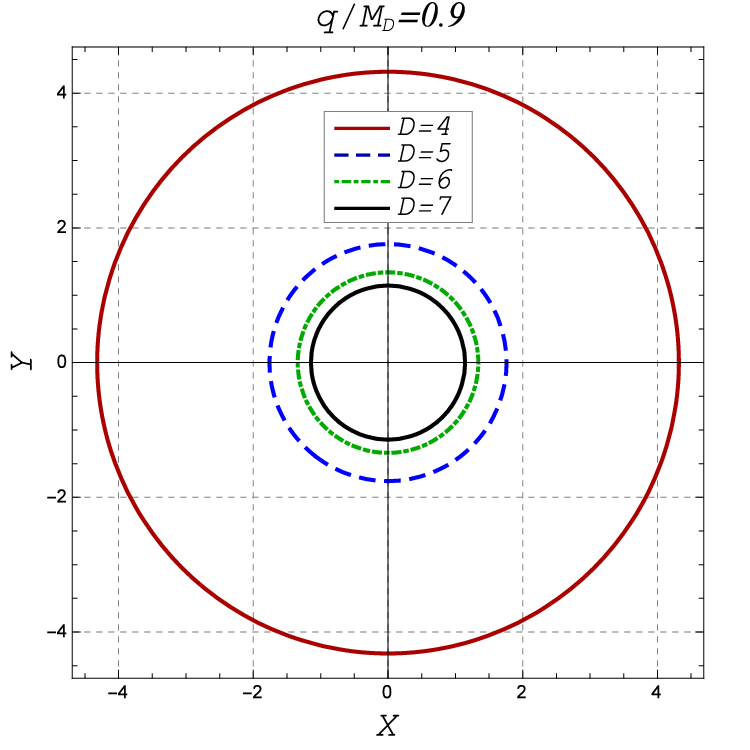}
    \caption{Plot showing the variation of the black hole shadow with the increasing number of spacetime dimensions $D$. }
    \label{sfig}
\end{figure}
where $\mathcal{K}$ is the separable constant usually known as Carter constant and first introduced by Carter in \cite{Carter:1968rr}. We get the complete null geodesics equations for the TRN black hole by substituting Eqs.~(\ref{p1})-(\ref{p2}) in Eqs.~(\ref{sr}) and (\ref{stheta}) in the given form
\begin{eqnarray}
&&\frac{dt}{d\tau}=\frac{\mathcal{E}}{1-{\frac{\mu}{r^{D-3}}+\frac{\nu^2}{r^{2(D-3)}}}}, \label{tp}\\	&&\frac{d\phi}{d\tau}=\frac{{L}}{r^2\displaystyle\prod\limits_{i=1}^{D-3}\sin^2\theta_i}, \label{p}\\
&&r^2 \frac{dr}{d\tau}=\pm\sqrt{\mathcal{R}},\label{r}\\
&&r^2\sum_{i=1}^{D-3}\prod_{n=1}^{i-1}\sin{\theta_i}\frac{d\theta_i}{d\tau}=\pm\sqrt{\Theta_i}, \label{th}
\end{eqnarray}
where the  parameter $\mathcal{R}(r)$ and $\Theta_{i}(\theta_{i})$ are the functions of radial and angular coordinates respectively, and take the following form
\begin{eqnarray}
&&\mathcal{R}(r)= \mathcal{E}^2r^4-r^2{\left[1-{\frac{\mu}{r^{D-3}}+\frac{\nu^2}{r^{2(D-3)}}}\right]}\left[\mathcal{K}+{{L}}^2\right],\\\label{R}
&&\Theta_i(\theta_i)=\mathcal{K}-\prod_{i=1}^{D-3}{{L}}^2\cot^2\theta_i.
\end{eqnarray}
The above equations from (\ref{tp}) to (\ref{th})  describe the motion of photons in the context of the TRN black hole. Now we define two parameters in terms of constants $\mathcal{E}$ and ${L}$ as: $\xi = {L}/\mathcal{E}$ and $\eta = {\mathcal{K}}/{\mathcal{E}^2}$. These impact parameters characterize the motion of photons around the black hole. 
The incoming photons towards the black hole may have two possibilities: they may fall inside the black hole or form bound orbits. The innermost unstable circular bound orbit around the black hole defines the black hole shadow boundary. The effective potential of the black hole can be used to analyze the circular bound orbits.  To get the expression of the  effective potential of the black hole, we redefine the radial equation of motion as
\begin{equation}
\left(\frac{d{r}}{d\tau}\right)^2+V_{eff}(r)=0.
\end{equation}
Here,  $V_{eff}$ is the effective potential of the black hole and can be expressed in terms of impact parameters as 
\begin{equation}
V_{eff}={\mathcal{E}^2}\left[\frac{1}{r^2}\left(1-{\frac{\mu}{r^{D-3}}+\frac{\nu^2}{r^{2(D-3)}}}\right)(\mathcal{\eta}+{\xi}^2)- 1 \right].
     \label{vef}
\end{equation}
In the limit of $D\rightarrow 4$, this effective potential reduces for the RN black hole, and in the absence of the black hole charge, the effective potential recovers its shape for the Schwarzschild black hole \cite{Singh:2017vfr}. Our interest is to find the innermost unstable circular orbits by maximizing the effective potential, which follows a given equation
\begin{equation}
	V_{eff}=0 \quad   \mbox{and}   \quad \frac{\partial V_{eff}}{\partial r}=0. \label{vr} 
\end{equation}
By maximizing the effective potential of the black hole, we obtain numerical values for the critical radius $r_c$ and the impact parameters $\eta+\xi^2$ with the increasing number of spacetime dimensions $D$. We show the variation of $r_c$ and $\eta + \xi^2$ with $D$ in Table~(\ref{table1}). In four-dimensional spacetime, which is the case of Schwarzschild black hole ($q=0$), the critical radius of innermost unstable circular orbits is $3 M_D$ (cf. Table~\ref{table1}) while in the presence of charge parameter $q$, it further decreases from $3 M_D$ to $2 M_D$ with the increasing values of charge parameter $q$ from $0$ to $1$ $M_D$. As we go to the higher-dimensional spacetime, the value of critical radius $r_c$ decreases, and in five-dimensional spacetime, it approaches $1.2917 M_D$ with $q = 0$ and further decreases as we consistently increase the number of spacetime dimensions (cf. Table~\ref{table1}). The critical radius $r_c$ also decreases with the increasing values of charge parameter $q$ in higher-dimensional spacetime. In higher-dimensional spacetime, the effect of charge parameter $q$ on the critical radius of the innermost circular orbits is small as compared to the four-dimensional spacetime.

\section{Black hole evaporation of TRN black hole} \label{sect4}

In this section, we study the black hole evaporation process for the TRN black hole. The black hole evaporates and loses its mass so the black hole mass decreases with time. Here in our study, we consider the emitted particles to be massless.
Let us consider a $(D-1)$- spatial dimensional cavity  and the number of modes for this cavity is given by  \cite{Vos1989, 0510002}
\begin{equation}
dN=\frac{dx_1 dx_2 \ldots d x_{D-1}d p_1 d p_2\ldots d p_{D-1}}{h^{D-1}} =  V \frac{ A p^{D-2} d p}{h^{D-1}},
\end{equation}
where $h$ is the Planck constant and $V$ is the $(D-1)$-dimensional cavity volume and $A={(D-1)\pi^{\frac{D-1}{2}}}/{\Gamma(\frac{D+1}{2})}$, is the area of the $(D-1)$-dimensional sphere of unit radius. In four-dimensional spacetime, it becomes $4 \pi$. The emitted radiation has $(D-2)$ independent polarization so we multiply this factor by the above equation \cite{Crispino:2000jx}, and we obtain
\begin{equation}
dN(\omega)= V \frac{(D-2)A\omega^{D-2} d\omega}{(2\pi)^{D-1}}, \label{vol}
\end{equation}
where $\omega$ is the frequency of the emitted radiation. One can obtain the expression of $(D-1)$-dimensional photon gas energy by multiplying the photon energy $\hbar \omega$ and the Bose-Einstein factor $1/(e^{({\hbar \omega}/T)} -1)$ in Eq.~(\ref{vol}) as
\begin{equation}
U = V\frac{(D-2)A}{(2\pi)^{D-1}}\int_{0}^{\infty} \hbar \omega\frac{\omega^{D-2}}{e^{\frac{\hbar\omega}{T}}-1} d\omega = V \frac{(D-2)A T^{D}}{(2\pi)^{D-1} \hbar^{D-1}}\int_{0}^{\infty}\frac{x^{D-1}}{e^x-1} d x.
\end{equation}
The energy density is proportional to the $T^D$. The expression of the  Stefan-Boltzmann constant in higher-dimensional spacetime is given by
\begin{equation}
k_D=\frac{(D-2)A}{(2\pi)^{D-1} \hbar^{D-1}}\int_{0}^{\infty}\frac{x^{D-1}}{e^x-1} dx.
\end{equation}
In four-dimensional spacetime, the expression of the energy density takes the following form as
\begin{equation}
U=\frac{\pi^2}{15 {\hbar}^3}VT^4.
\end{equation}
In the geometrical optics approximation,  the emitted massless quanta particle will follow the path of the null geodesics \cite{Page:2015rxa}. The radial null geodesics equation of motion of the emitted particle from the black hole reads
\begin{equation}
    \left(\frac{dr}{d\tau}\right)^2 = E^2 - J^2\frac{\left(1-{\frac{\mu}{r^{D-3}}+\frac{\nu^2}{r^{2(D-3)}}}\right)}{r^2}
\end{equation}
where $E$ and $J$ are the energy and the angular momentum of the emitted quanta particle from the black hole. The condition for  emitted radiation to reach infinity rather than falling inside the black hole is
\begin{equation}
 \frac{E^2}{J^2} \geq  \frac{\left(1-{\frac{\mu}{r^{D-3}}+\frac{\nu^2}{r^{2(D-3)}}}\right)}{r^2}.
\end{equation}
The ratio of $J$ and $E$ is the impact parameter and the critical value of this impact parameter is given by \cite{Page:2015rxa}
\begin{equation}
    \xi_c = \frac{r_p}{\sqrt{\left(1-{\frac{\mu}{r_p^{D-3}}+\frac{\nu^2}{r_p^{2(D-3)}}}\right)}}.
\end{equation}
As we obtained the critical value of the impact parameter,  now according to the $D$-dimensional Stefan-Boltzmann law \cite{Landsberg:1989, Cardoso:2005cd}, the Hawking emission power is
\begin{equation}
    \frac{dM}{dt} = - \mathcal{C} \xi_{c}^{D-2} T^D,
\end{equation}
where $\mathcal{C} = (D-2) \pi^{\frac{D}{2}-1} \Omega_{D-2} \frac{\Gamma(D)}{\Gamma(\frac{D}{2})}\zeta(D)$. The black hole mass loss rate is dimension-dependent and as we move towards higher-dimensional spacetime the black hole mass decreases faster and the black hole evaporates rapidly.
\section{ {TRN black hole Shadow} } \label{sect5}
In this section, we estimate the geometrical radius of the photon orbits around the TRN black hole. The rotating black hole shadow in the higher-dimensional spacetime has an oblate shape due to the angular momentum of the black hole and spacetime dimensions $D$ \cite{Singh:2023ops}. The TRN is a nonrotating higher-dimensional class of black holes, so the observed photon orbits are circular. To visualize these photon orbits, we define celestial coordinates $X$ and $Y_{i}$. These celestial coordinates can be defined via
\begin{eqnarray}
&&X=\lim_{r_0\rightarrow\infty}\left(\frac{r_0P^{(\phi)}}{P^{(t)}}\right),\label{al}\\
&&Y_i=\lim_{r_0\rightarrow\infty}\left(\frac{r_0P^{(\theta_i)}}{P^{(t)}}\right) , \quad \:\:\:  i=1, . . . ,D-3,\label{be}
\end{eqnarray}
where $P^{(\phi)}$, $P^{(t)}$ and  $P^{(\theta_{i})}$ are the vi-tetrad component of momentum and $r_0$ is the distance between the black hole and the far observer \cite{Chandrasekhar:1992}. 
\begin{figure}
    \centering
    \includegraphics[scale = 0.7]{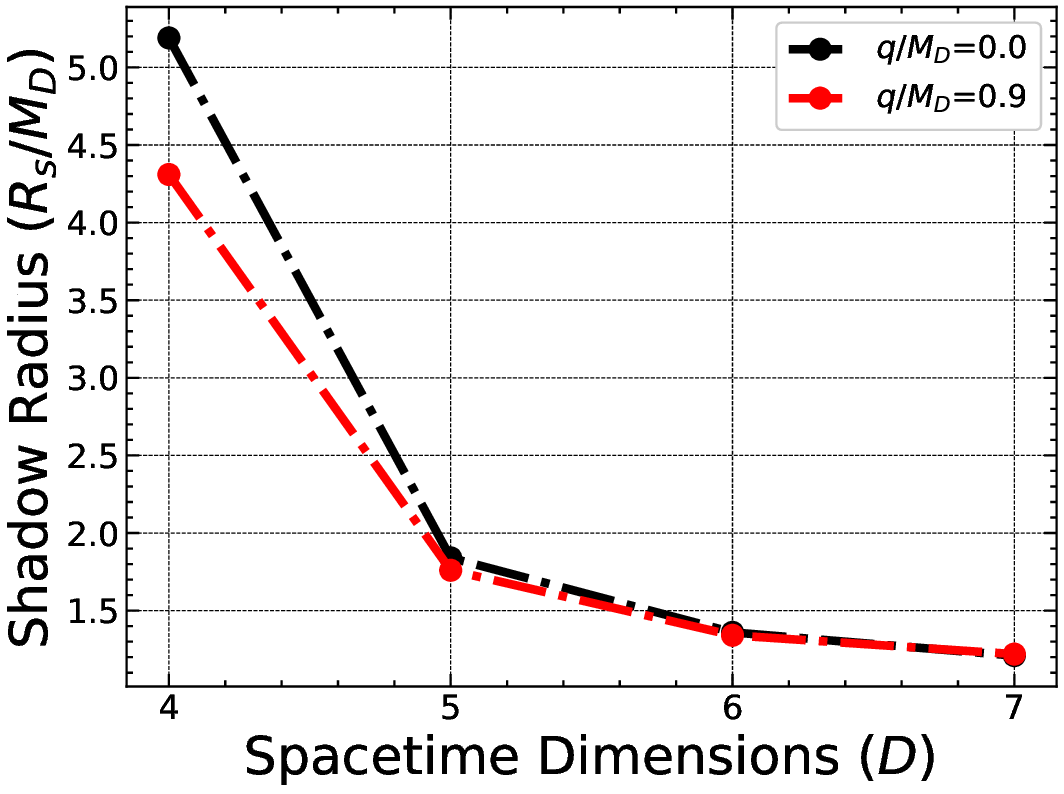}
    \caption{Plot showing the variation of shadow radius with spacetime dimensions $D$. }
    \label{figure2}
\end{figure}
Substituting  Eqs.~(\ref{p1})-(\ref{p2}) and  (\ref{tp})-(\ref{th}) in the above celestial coordinate Eqs.~(\ref{al})-(\ref{be}), we find celestial coordinates in terms of impact parameters and inclination angle $\theta_i$, which reads
 \begin{eqnarray}
&&X=\frac{-\xi}{\displaystyle\prod_{i=1}^{D-3}\sin\theta_i},\label{a}\\
&&Y_i=\pm\sqrt{\eta-\xi^2\prod_{i=1}^{D-3}\cot^2\theta_i}.\label{b}
\end{eqnarray}
For simplicity, we choose equatorial plane where $\theta_{i}=\pi/2$ and our equations of celestial coordinate reduces to simpler form
\begin{equation}\label{alpha-beta1}
X=-\xi, \;\;\;  Y=\pm\sqrt{\eta}, 
\end{equation}
and the above Eq.~(\ref{alpha-beta1}) must follow the condition  
\begin{eqnarray}
X^2+Y^2  = \eta +\xi ^2.  \label{n}
\end{eqnarray}
The contour plot of the above equation traces the photon orbits around the TRN black hole.  In four-dimensional spacetime, the above Eq.~(\ref{n}) reduces to the  RN black hole, and in the absence of charge parameter $q\rightarrow 0$, it recovers the results for Schwarzschild black holes in higher-dimensional spacetime \cite{Singh:2017vfr}. We trace several contour plots of black hole shadow with the variation of spacetime dimension $D$ and charge parameter $q$ (cf. Fig.~\ref{figure1}).  In higher-dimensional spacetime, the black hole shadow rapidly decreases (cf. Fig.~\ref{figure1} and Fig.~\ref{sfig}).  In  Fig.~(\ref{figure1}), the inner dashed black circular ring shows the shapes of the TRN black hole and the outer orange circular ring shows the shapes of Schwarzschild-Tangherlini black holes. In Fig.~(\ref{figure2}), we plot the shadow radius as a function of the spacetime dimensions $D$. The shadow radius $R_s$ decreases monotonically with increasing spacetime dimensions.  In four-dimensional spacetime, the radius of the black hole shadow $R_s$  varies from ($5.19$ to $4.31$ $M_D$) for ($q = 0$ to $0.9$ $M_D$). Similarly, for seven-dimensional spacetime, the shadow radius varies from ($1.15$ to $1.14$ $M_D$) for ($q = 0$ to $0.9$ $M_D$). From Fig.~(\ref{figure1}), (\ref{sfig}) and (\ref{figure2}), it is noticeable that the impact of the charge parameter on the size of the black hole shadow decreases as we move towards higher-dimensional spacetime. The effect of spacetime dimensions becomes more dominant than the effect of charge as we move toward higher-dimensional spacetime.
\section{Consistency with M87 and Sgr A$^{*}$ black hole observations}  \label{sect6}
The recent observations from Very Long Baseline Interferometry provide new insights into gravity in strong gravitational regimes. Near the horizon of the supermassive black hole, the deflection angle is unbound, and the photons are trapped into the indefinite orbits. These indefinite orbits are called photon shells. In this section, we find the consistency of the higher-dimensional charged black hole with the astrophysical black hole M87 and Sgr A$^{*}$. Here we calculate the angular diameter of the charged black hole shadow in the higher-dimensional spacetime with the observed data of astrophysical black holes from EHT collaborators. The angular diameter of the black hole shadow can be calculated using the following equation \cite{Kumar:2018ple}
\begin{equation}    \label{angular}
    \theta_{D} = \frac{2}{r_{0}} \times \sqrt{\frac{A_{0}}{\pi}}     
\end{equation}
where $A_{0}$ is the area of the black hole shadow and $r_{0}$ is the observer's distance from the black hole shadow. We use Eq.~(\ref{angular}) to estimate the angular diameter of the higher-dimensional charged black hole shadow.
\begin{figure}
    \centering
    \includegraphics[scale =0.9]{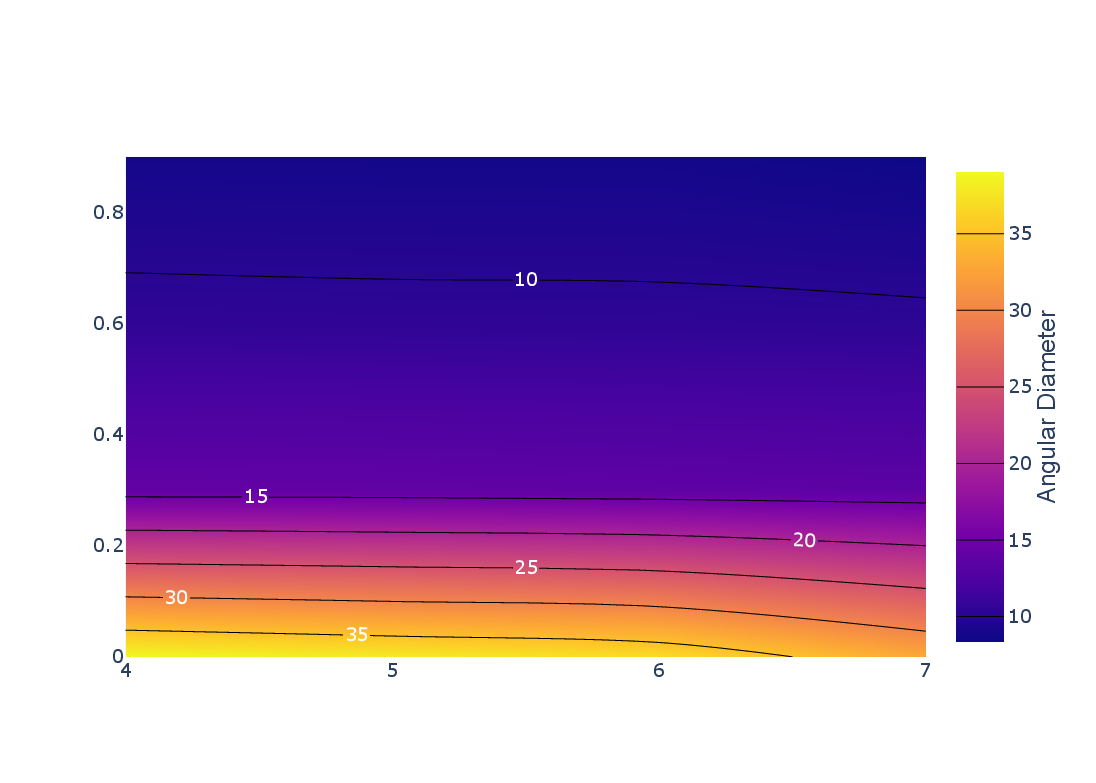}
    \caption{ Variation of angular diameter $\theta_D$ with the spacetime dimension $D$ and charge $q$ with M87 black hole observations. }
    \label{figer3}
\end{figure}
\subsection{M87 black hole }
In April 2019, EHT published the first image of the M87 black hole with its astrophysical properties in \cite{EventHorizonTelescope:2019dse, EventHorizonTelescope:2019uob, EventHorizonTelescope:2019jan, EventHorizonTelescope:2019ths, EventHorizonTelescope:2019pgp, EventHorizonTelescope:2019ggy}. The M87 black hole has mass $6.5 \times 10^{9} M_{\odot}$ and is $16.8$ $Mps$ distant from the earth. We analytically estimate the angular diameter of the higher-dimensional charged black hole shadow with the M87 black hole observations. The estimated angular diameter in four-dimensional spacetime varies from $\theta_D \approx$ $(39$ to $33)$ $\mu as$ for $q \rightarrow$ $(0$ to $0.9 )$ $M_D$ . The angular diameter of the black hole shadow decreases monotonically in the higher-dimensional spacetime. In $D=7$, the angular diameter varies from $\theta_D \approx$ $(13$ to $7)$ $\mu as$ for $q \rightarrow$ $(0$ to $0.9)$ $M_D$ (cf. Fig.~\ref{figer3}).
\begin{figure}
    \centering
    \includegraphics[scale =0.9]{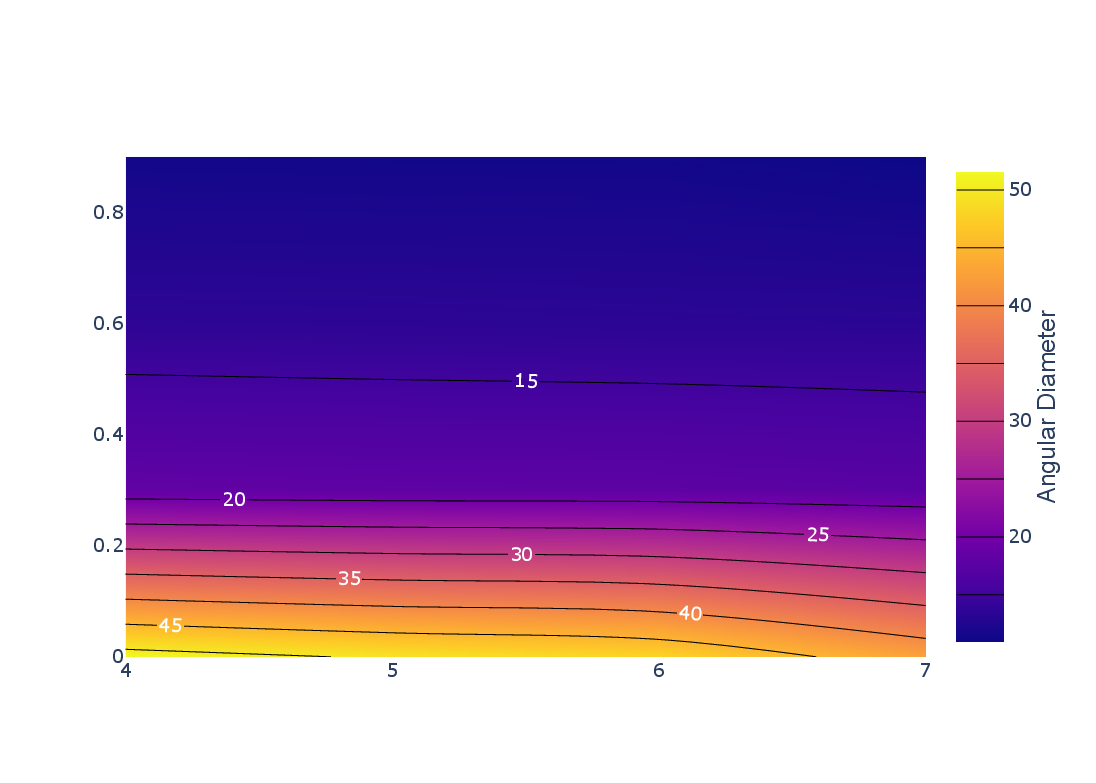}
    \caption{Variation of angular diameter $\theta_D$ with the spacetime dimension $D$ and charge $q$ with Sgr A$^*$ black hole observations. }
    \label{figure4}
\end{figure}
\begin{figure}
    \centering
    \includegraphics[scale = 0.64]{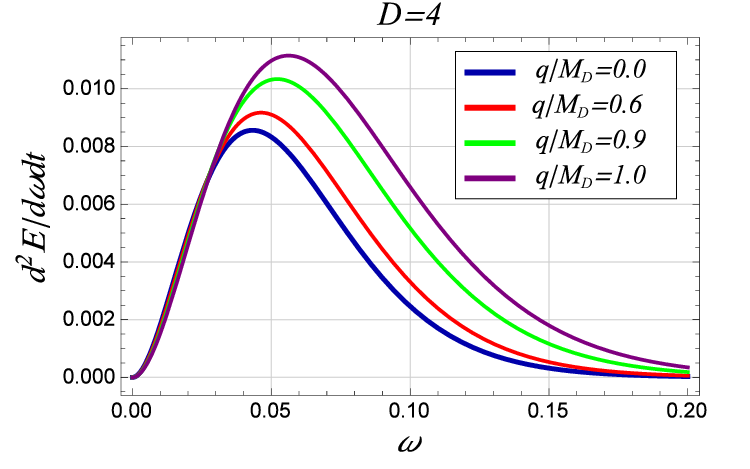}
     \includegraphics[scale = 0.64]{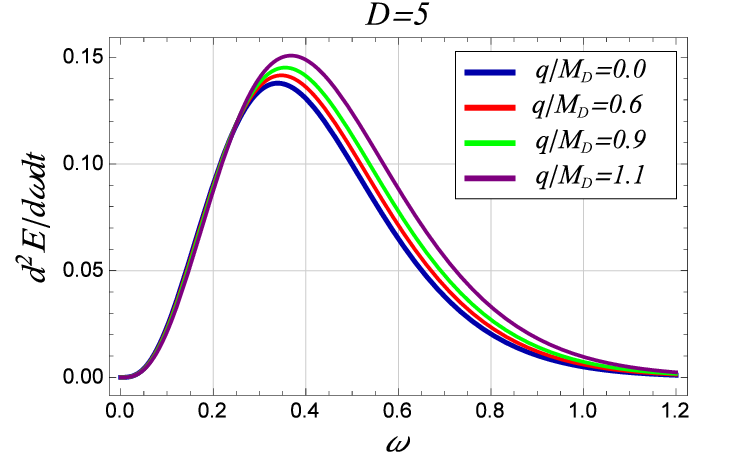}\\
     \includegraphics[scale = 0.64]{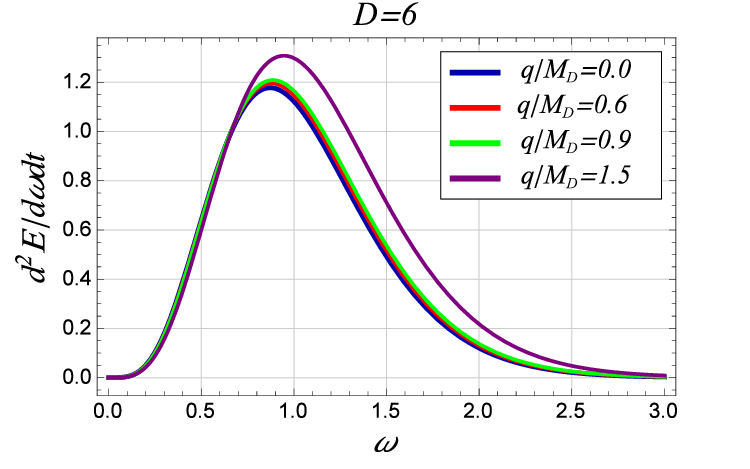}
     \includegraphics[scale = 0.64]{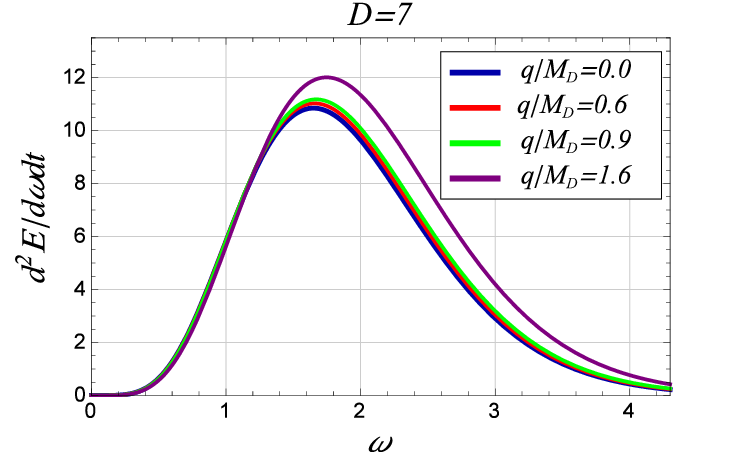}
    \caption{Variation of the energy emission rate with spacetime dimensions and black hole charge.  }
    \label{f3}
\end{figure}
\subsection{Sgr A$^*$ black hole}
The observed results  of Sgr A$^*$ black hole was published by EHT collaborators in \cite{EventHorizonTelescope:2022wkp, EventHorizonTelescope:2022apq, EventHorizonTelescope:2022wok, EventHorizonTelescope:2022exc, EventHorizonTelescope:2022urf, EventHorizonTelescope:2022xqj}. The Sgr A$^*$ black hole is situated at the center of our galaxy Milky Way. The observed mass of this black hole is $4 \times 10^{6}$ $M_{\odot}$ and its $8 kps$ distance far from the Earth. In four-dimensional spacetime, the estimated angular diameter of Sgr A$^*$ black hole is  $(50$ to $40)$ $\mu as$ for $q \rightarrow$ $(0$  to $0.9)$ $M_D$ and as we go to the higher-dimensional spacetime the angular diameter of the black hole shadow decreases. For $D=7$,  the angular diameter take values from $\theta_D$ $\approx$ $(11$ to $9 )$ $\mu as$ for $q \rightarrow$ $(0$ to $0.9)$ $M_D$ (cf. Fig.~\ref{figure4}).

The angular diameter of the charged black hole shadow decreases in the higher-dimensional spacetime. The TRN black hole shadow in the higher-dimensional spacetime appears smaller in comparison with four-dimensional spacetime.
\section{Energy emission rate}\label{sect7}
In our previous sections (\ref{sect4}) and (\ref{sect5}), we have discussed that the black hole loses its mass in higher-dimensional spacetime, and the size of the black hole shadow monotonically decreases with the increasing spacetime dimensions. In that sequence, we aim to determine how the energy emission rate varies as the number of spacetime dimensions increases. The emission rate of the black hole in higher-dimensional spacetime is given by \cite{Singh:2017vfr}
\begin{equation}
    \frac{d^2E}{d\omega dt} = \frac{2 \pi^2 \sigma_{lim}}{ \left(\exp({\frac{\omega}{T}})-1 \right)}  \omega^{(D-1)}
\end{equation}
where $\omega$ is the frequency of the emitted radiation, $\sigma_{lim}$ is the limiting constant value. 
The shadow of the black hole corresponds to the high energy absorption cross-section for a far-distant observer \cite{Misner:1973prb}. The absorption cross sections oscillate to a limiting constant value  $\sigma_{lim}$ for a spherically symmetric black hole and can be estimated via the area of the photon sphere \cite{ Wei:2013kza, Decanini:2011xi}.  In higher-dimensional spacetime, the $\sigma_{lim}$ can be defined as
\begin{equation} 
    \sigma_{lim} \approx \frac{\pi^{\frac{D-2}{2}} R_{s}^{(D-2)}}{\Gamma(\frac{D}{2})},
\end{equation}
where $R_s$ is the black hole shadow radius. The limiting constant value is reduced to $\pi R_s^2$ in four-dimensional spacetime. The expression of energy emission rate in higher dimensional spacetime is expressed by
\begin{equation}
    \frac{d^2E}{d\omega dt} = \frac{2 \pi^{(\frac{D+2}{2})} (\omega R_s)^{(D-2)} }{(e^{\omega/T} - 1)\Gamma(D/2)} \omega.
\end{equation}
The variation of the energy emission rate of the charged black hole with the increasing number of spacetime dimensions has been shown in Fig.~(\ref{f3}). In the first plot of Fig.~(\ref{f3}), we vary the charge of the black hole in the four-dimensional spacetime. As the charge parameter $q$ increases, the rate of energy emission also increases. The second plot of Fig.~(\ref{f3}) shows the emission rate in five-dimensional spacetime. Here, we can see the energy emission rate increases compared to the four-dimensional spacetime. The third and the fourth plots of Fig.~(\ref{f3}) show the variation of energy emission rate in sixth and seventh-dimensional spacetime. The emission rate increases as we move to higher-dimensional spacetime. However, the effect of the charge on the emission rate is effectively small compared to four-dimensional spacetime.
\section{Concluding Remarks}\label{sect8}
Our study reveals that the shadow of a charged black hole decreases in size when observed in spacetimes with a higher number of dimensions. The black hole evaporation process occurs very fast in the higher dimensional spacetime. Due to this evaporation process, the black hole radiates more and loses its mass \cite{Kanti:2014vsa}. The horizon of the black hole shrinks and the effective size of the black hole shadow decreases. The key results of our study are given below
\begin{itemize}
    \item The effective potential of the higher-dimensional charged black hole has been originally derived, and the numerical study of the critical radius of photon orbits has been done with the effect of spacetime dimensions.
    \item With the definition of the impact parameters, the geometry of the photon orbits has been studied for the increasing number of spacetime dimensions.
    \item We derived the equation of the rate of mass loss and concluded that the higher-dimensional charged black hole evaporates rapidly.
    \item The size of the black hole shadow decreases with spacetime dimensions $D$ and we have shown several plots of black hole shadow for the increasing number of spacetime with the effect of charge.
    \item The angular diameter of the charged black hole shadow has been estimated with the given data of M87 and Sgr A$^{*}$ black holes.
    \item The angular diameter of the charged black hole shadow rapidly decreases in higher-dimensional spacetime in comparison to the Schwarzschild-Tangherlini black hole which emphasizes that the higher-dimensional charged black hole shadow appears much smaller in comparison to the higher-dimensional Schwarzschild black hole \cite{Singh:2017vfr}.
\end{itemize}
It is now clear from EHT observations that black holes exist in our Universe, however, we are still missing the signature of higher-dimensional astrophysical black holes. Very soon, with the help of the Next Generation Event Horizon Telescope, we may get more images of other black holes in the Universe, where we can test higher-dimensional black hole models. In that case, our study might be helpful. 

\textbf{Acknowledgement}
The author would like to express gratitude to Prof. Sushant G. Ghosh for his unwavering support.
\\
\textbf{Credit authorship contribution statement}\\
Balendra Pratap Singh: Conceptualization, Methodology, Software, Visualization, Investigation, Writing, Review and Editing.
\\
\textbf{Declaration of competing interest:} \\
The author declares that they have no known competing financial interests or personal relationships that could have appeared
to influence the work reported in this paper.
\\
\textbf{Data availability:}\\
No data has been used for the research described in the article.

\end{document}